\documentclass[aps,pre,twocolumn,notitlepage, showpacs,superscriptaddress,10pt,floatfix]{revtex4-2}

\usepackage{graphicx}
\usepackage{grffile}
\usepackage{amsthm}
\usepackage{epsfig}
\usepackage{amsmath,amssymb,amsfonts,mathtools,bbm,MnSymbol,mathrsfs,xfrac}
\usepackage{longtable}
\usepackage{tabularx}
\usepackage{multirow}
\usepackage{makecell}
\usepackage{colortbl}
\usepackage[table,svgnames,xcdraw]{xcolor}
\usepackage{array, graphicx}
\usepackage[breaklinks=true,colorlinks=true,linkcolor=blue,urlcolor=blue,citecolor=blue]{hyperref}
\usepackage{comment}
\usepackage{color}
\usepackage{hhline}
\usepackage[makeroom]{cancel}

\usepackage{tikz}
\usetikzlibrary{decorations.pathreplacing,calligraphy}
\usepackage{pgfplots}
\usepackage{tikz-3dplot}
\tdplotsetmaincoords{60}{115}
\pgfplotsset{compat=newest}

\usepackage{soul}

\renewcommand{\[}{\begin{equation}}
\renewcommand{\]}{\end{equation}}

\newcommand{\ket}[1]{|#1\rangle}
\newcommand{\bra}[1]{\langle#1|}

\newcommand{\pro}[2]{|#1\rangle\!\langle#2|}
\newcommand{\mean}[1]{\left\langle#1\right\rangle}

\newcommand{\h}{{\hat{H}}}

\newcommand{\bs}{\mathbf{s}}

\definecolor{mygray}{gray}{0.6}

\theoremstyle{definition}

\definecolor{dfcol}{cmyk}{1, 0.2108, 0.13, 0.3}
\newcommand{\df}[1]{\ifthenelse{\boolean{}}{\textcolor{dfcol}{[{\bf DF}: #1]}}{}}

\usepackage{physics}
\usepackage[page]{appendix}

\pdfoutput=1

\begin{document}


\title{Boltzmann Sampling by Diabatic Quantum Annealing}
\author{Ju-Yeon Gyhm}
\thanks{These authors contributed equally to this work.}
\affiliation{%
Department of Physics and Astronomy \& Center for Theoretical Physics, Seoul National University, Seoul 08826, Korea
}%

\author{Gilhan Kim}
\thanks{These authors contributed equally to this work.}

\affiliation{%
Department of Physics and Astronomy \& Center for Theoretical Physics, Seoul National University, Seoul 08826, Korea
}%
\thanks{Present address: Department of Statistics and Data Science, Yonsei University, Seoul 03722, Korea}

\author{Hyukjoon Kwon}
\affiliation{%
School of Computational Sciences, Korea Institute for Advanced Study, Seoul 02455, Korea
}%

\author{Yongjoo Baek}
\email{y.baek@snu.ac.kr}

\affiliation{%
Department of Physics and Astronomy \& Center for Theoretical Physics, Seoul National University, Seoul 08826, Korea
}%

\date{\today}
\begin{abstract}
Boltzmann sampling is a central component of many computational frameworks, including numerous algorithms in machine learning. Although quantum annealers have been investigated as potential fast Boltzmann samplers, their dependence on environmental noise makes precise control of the effective temperature difficult, introducing uncertainty into the sampling process. As an alternative, we propose diabatic quantum annealing---a faster, purely unitary process---as a controllable Boltzmann sampler in which the effective temperature is determined by the annealing rate. Using the ferromagnetic Ising model and the Sherrington--Kirkpatrick model as test cases, we demonstrate that this method achieves rapid and accurate sampling in the high-temperature regime.
\end{abstract}

\maketitle

\section{Introduction} \label{sec:intro}


\textit{Boltzmann sampling} plays a central role in a wide range of numerical studies. Beyond its conventional use for estimating statistical properties at a fixed (effective) temperature, it serves as a key component of energy-based machine learning models, including Boltzmann machines~\cite{Hinton1983}, restricted Boltzmann machines~\cite{Smolensky1986,Hinton2002}, and deep belief networks~\cite{Hinton2006}. Its typical implementation relies on Markov chain Monte Carlo (MCMC) methods, which involve local stochastic updates~\cite{Newman1999} and are prone to critical slowing down or trapping in glassy energy landscapes. Classical techniques like simulated annealing~\cite{KirkpatrickScience1983} and parallel tempering~\cite{SwendsenPRL1986} can alleviate these issues to some extent, but the task remains NP-hard in general~\cite{Barahona1982}, making it a major computational bottleneck.

\begin{figure}
\includegraphics[width=\columnwidth]{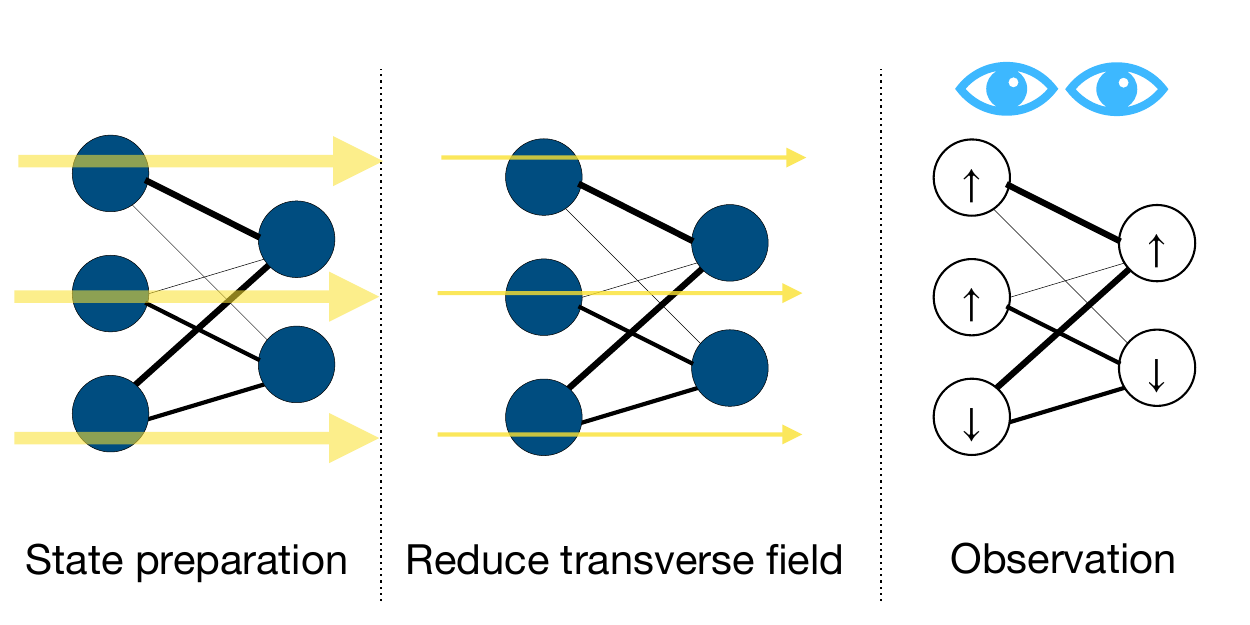}
\caption{\label{fig:DQAmain} Schematic of quantum annealing. The optimization problem is encoded in the fields and couplings of an Ising Hamiltonian. Initially, the system is prepared in the ground state under a strong transverse field. This field is then gradually reduced, leaving only the problem Hamiltonian. In an ideal adiabatic and unitary process, the system remains in the ground state throughout, yielding the optimal solution. However, imperfections in the annealing schedule can lead to transitions into excited states, which may be observed in the final outcome.}
\end{figure}

Meanwhile, \textit{quantum annealing} (QA)~\cite{Kadowaki1998,BrookeScience1999,FarhiScience2001,RajakPTRSA2023} was originally proposed as an optimization technique. As illustrated in Fig.~\ref{fig:DQAmain}, QA begins with a strong transverse field that prepares the system in its ground state. This field is then slowly turned off, leaving only the problem Hamiltonian. If the evolution is fully unitary and adiabatic, the quantum adiabatic theorem~\cite{kato1950adiabatic} ensures that the system ends up in the ground state of the final Hamiltonian. However, even in a state-of-the-art implementation of quantum annealing (the D-Wave platform), the dephasing time is on the order of $10~\mu \mathrm{s}$~\cite{dwave}, which is comparable to the typical time span of a standard annealing schedule~\cite{dwave-annealing-time}. As a result, real-world QA is subject to environmental noise, and final states often include excitations above the ground state.

These \textit{imperfections} of quantum annealers have opened up their use as Boltzmann samplers instead of optimizers. Two different approaches have emerged in the literature. First, one can collect low-lying states from the quantum annealer and use them to estimate ensemble averages via manually assigned Boltzmann weights~\cite{SandtSciRep2023} or as seeds to enhance classical Boltzmann samplers~\cite{ScrivaSciPost2023,Arai2025}. This method has the advantage of being agnostic to the final-state distribution, but extra efforts are required to capture high-energy configurations with low probability.

The second approach is based on the observation that quantum annealers can produce distributions closely matching the classical Boltzmann distribution~\cite{Bian2010,PerdomoOrtizSciRep2016,BenedettiPRA2016}. This is attributed to fast thermalization occurring in the early phase of the annealing schedule, when transitions between states are frequent due to the strong transverse field. Such transitions become rarer as the transverse field weakens, eventually freezing the state populations. If this freeze-out event occurs rapidly in a short period of time, and if the transverse field has already become very weak by then, the final state follows the Boltzmann distribution whose effective temperature is determined by where the freeze-out point falls in the annealing schedule~\cite{AminPRA2015}. This has led to proposals of using quantum annealers as fast Boltzmann samplers, with applications in supervised learning~\cite{Bian2010,Adachi2015,Dorband2015,AminPRX2018,LiuEntropy2018,LiQST2020,Caldeira2020,Krzysztof2021}, unsupervised learning~\cite{Denil2011,Dumoulin2014,BenedettiPRA2016,AminPRX2018,BenedettiPRX2017,LiQST2020,Sleeman2020,Dixit2022,MoroCommPhys2023}, and reinforcement learning~\cite{CrawfordQIC2018}.

However, using this approach, practical limitations hinder precise control of the effective temperature. Since the Hamiltonian implemented by the D-Wave platform may deviate from the original one~\cite{dwave-error,Pochart2022}, the freeze-out point can vary from one instance to another, inducing variations in the effective temperature~\cite{BenedettiPRA2016,RaymondICT2016}. Moreover, depending on the problem, the transverse field may not be sufficiently weak at the freeze-out point, in which case the final state is not well described by the classical Boltzmann distribution~\cite{MarshallPRAppl2017}. Strategies such as pausing the annealing schedule~\cite{MarshallPRAppl2019} or correcting for spurious couplings~\cite{VuffrayPRX2022,NelsonPRAppl2022} can mitigate some of these issues. Yet, as long as the system interacts with the environment, tuning the effective temperature remains inherently difficult.

In this paper, we propose a third approach: diabatic quantum annealing (DQA). In DQA, the annealing is carried out over a much shorter time scale---on the order of nanoseconds, which is feasible on current D-Wave platforms---to ensure that the dynamics remain unitary. If the annealing rate $\alpha$ is infinite, the transverse field is instantaneously quenched to zero, which yields the infinite-temperature Boltzmann distribution (whose inverse temperature is $\beta = 0$). We show that, when $\alpha$ is very large but finite, there exists an approximate relation between $\alpha$ and $\beta$ that depends only on how many Ising spins are involved in each term of the Hamiltonian. Using this relation and an appropriately rescaled Hamiltonian, in the high-temperature regime, we can fix the effective temperature of the Boltzmann distribution only by controlling the annealing rate. While conventional MCMC methods do not suffer from critical slowing down and dynamic arrest in this regime, DQA can still offers significant advantages via its fast dynamics and intrinsic parallelism---a single DQA run generates statistically independent samples in nanoseconds, regardless of the system size. We demonstrate the viability of DQA through proof-of-concept simulations on three representative systems: the ferromagnetic Ising model on an all-to-all network, the ferromagnetic Ising model on a two-dimensional square lattice, and the Sherrington–Kirkpatrick model.

The rest of the paper is organized as follows. In Sec.~\ref{sec:DQA}, we describe our theory that matches annealing rate to effective temperature. In Sec.~\ref{sec:sampling_ferro}, we check how well our proposed method reproduces the Boltzmann statistics of the representative systems. In Sec.~\ref{sec:theory}, we show the details of our analytical derivation. Finally, we summarize our findings and conclude in Sec.~\ref{sec:conclusion}.

\section{Diabatic Quantum Annealing} \label{sec:DQA}

In this section, we present a recipe for approximating the Boltzmann statistics associated with an Ising Hamiltonian via DQA. First, we assume that the problem of interest is described by the energy function
\begin{align} \label{eq:Es}
E_\bs = -\sum_{1\leq i\leq N} h_i s_i - \sum_{1\leq i<j\leq N} J_{ij} s_i s_j\;,
\end{align}
where each Ising spin $s_i$ is either $+1$ or $-1$. At inverse temperature $\beta$, the associated Boltzmann distribution is given by $P_\mathrm{B}(\bs) \propto \mathrm{e}^{-\beta E_\bs}$, whose high-temperature expansion can be written as
\begin{align} \label{eq:PB}
P_\mathrm{B}(\bs) = \frac{1}{2^N} \left[1 - \beta E_\bs + \mathcal{O}(\beta^2 E_\bs^2)\right]\;.
\end{align}
We note that this probability distribution satisfies the normalization condition to order $\beta$ for any system described by Eq.~\eqref{eq:Es} since $\sum_\bs E_\bs = 0$ holds for any choice of $h_i$'s and $J_{ij}$'s.

Our goal is to design a QA process whose final state correctly reproduces the above distribution to the order $\beta E_\bs$. Toward this end, we set up the time-dependent Hamiltonian
\begin{align} \label{eq:diabatic}
\h(t)= A(t)\h_x+B(t)\h_z\;, 
\end{align}
where the component Hamiltonians are given by
\begin{align} \label{eq:HzHx}
\h_x &= \sum_{i=1}^N \hat{\sigma}^x_i\;, \nonumber\\
\h_z &= -\sum_{1\leq i\leq N} h_i \hat{\sigma}_i^z - \frac{1}{c_2}\sum_{1\leq i<j\leq N} J_{ij}\hat{\sigma}_i^z\hat{\sigma}_j^z\;.
\end{align}
Here, $\hat{\sigma}^z_i$ and $\hat{\sigma}^x_i$ are the Pauli matrices, and $c_2$ is a rescaling factor whose value will be given shortly. In the initial state ($t = 0$), the magnitudes of $\h_x$ and $\h_z$ satisfy $A(0)\gg B(0) \ge 0$, so that the system is prepared in the ground state of $\h_x$ denoted as
\begin{align}
\ket{\psi_\mathrm{i}}=\bigotimes_{i=1}^N\ket{X-}_i\;,
\end{align}
with $\ket{X-}_i$ standing for the eigenstate of $\hat{\sigma}_i^x$ satisfying $\hat{\sigma}_i^x\ket{X-}_i=-\ket{X-}_i$. As the annealing proceeds, the values of $A(t)$ and $B(t)$ vary over time, so that $0 \le A(\tau)\ll B(\tau)$ holds in the end. The final state $\ket{\psi_\mathrm{f}}$, achieved at the end of the annealing process, is given by
\begin{align}\label{eq:unitary_evol}
    \ket{\psi_\mathrm{f}}=\hat{\mathcal{T}} \exp \left[-i \int_0^\tau  dt\,\h(t)\right] \ket{\psi_\mathrm{i}}\;,
\end{align}
where $\hat{\mathcal{T}}$ is the time-ordering operator, and we used the atomic units that fix $\hbar = 1$. If $A(t)$ and $B(t)$ change at a rate much smaller than the minimum energy gap of $\h_z$, the adiabatic theorem guarantees that $\ket{\psi_\mathrm{f}}$ is the ground state of $\h_z$. However, if $A(t)$ or $B(t)$ changes at a rate much greater than or comparable to the minimum energy gap of $\h_z$, $\ket{\psi_\mathrm{f}}$ will be distributed over excited states as well as the ground state. This allows us to define the projected probability distribution
\begin{align}
    P_\mathrm{Q}(\bs) = |\langle\bs|{\psi_\mathrm{f}}\rangle|^2\;.
\end{align}

\begin{figure*}
\includegraphics[width=\textwidth]{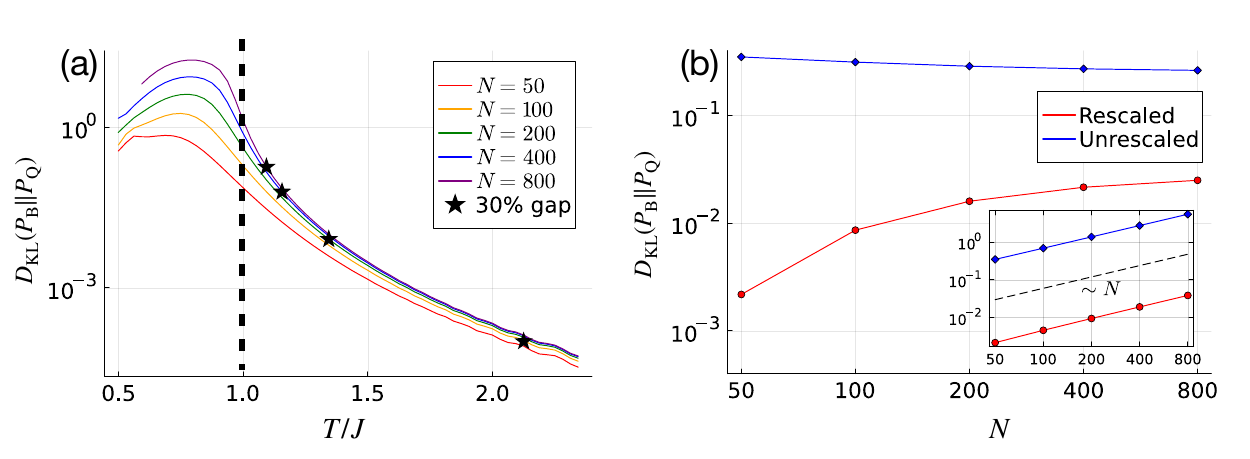}
\caption{\label{fig:kld} Kullback--Leibler (KL) divergence of the Boltzmann statistics $P_\mathrm{B}$ from the distribution $P_\mathrm{Q}$ sampled via DQA of the infinite-range Ising model. (a) For the case without the magnetic field ($h = 0$), the values of $T$ below which the KL divergence obtained at system size $N$ becomes $30\%$ larger than that obtained at system size $N/2$ are marked by the star symbols ($\bigstar$). The horizontal locations of the stars approach $T_\mathrm{c}/J$ (marked by the vertical dashed line) as $N$ grows, which suggests that, in the paramagnetic phase, the KL divergence saturates to a finite value in the thermodynamic limit. (b) For the case where $T/J = 2$ and $h N^{1/2}/J = \sqrt{50}/2$, the KL divergence converges to a finite value as $N$ goes to infinity. Inset: If we fix $h = 1/2$, the KL divergence grows linearly with $N$. In both cases, $P_\mathrm{Q}$ is much closer to $P_\mathrm{B}$ if we rescale the two-body interaction terms in the Hamiltonian by a factor of $1/\sqrt{2}$.}
\end{figure*}

In Sec.~\ref{sec:theory}, using the weak coupling assumption that allows us to truncate the Dyson series expansion of Eq.~\eqref{eq:unitary_evol} at first order, we prove that by fixing the rescaling factor in Eq.~\eqref{eq:HzHx} at
\begin{align}\label{eq:c_2_theory}
    c_2 = \frac{\int_0^\tau dt B(t)\sin\left[4\int_t^\tau ds A(s)\right]}{\int_0^\tau dt B(t)\sin\left[2\int_t^\tau ds A(s)\right]}\;,
\end{align}
the projected distribution $P_\mathrm{Q}$ reproduces the Boltzmann distribution $P_\mathrm{B}$ to the order $\beta E_\mathrm{s}$ with
\begin{align} \label{eq:theoretical_beta}
    \beta = 2\int_0^\tau dt B(t)\sin\left[2\int_t^\tau\!\! ds A(s)\right]\;.
\end{align}
These results are valid for any $A(t)$ and $B(t)$ satisfying the boundary conditions at the initial and final states. In Table~\ref{table:D-Wave}, we list the values of $\beta$ and $c_2$ achieved by the fast annealing schedule (which is distinct from the standard annealing schedule) offered by different versions of the D-Wave Advantage2 systems as the annealing time $\tau$ varies from $5$ to $10$~ns. We note that the listed values of $\beta$ and $c_2$ are obtained by plugging the schedules of $A(t)$ and $B(t)$ specified in \cite{dwave-schedule} into Eqs.~\eqref{eq:c_2_theory} and \eqref{eq:theoretical_beta}.
\begin{table}[h]
\centering

\begin{tabular}{c | c c c c c c}
\hline
\multicolumn{7}{c}{$\beta$} \\ 
\hhline{=======}
Version & 5~ns & 6~ns & 7~ns & 8~ns & 9~ns & 10~ns \\
\hline                                                                          System1.13 & 1.2596 & 1.5293 & 1.8383 & 2.1689 & 2.4507 & 2.8348 \\
System2.1 & 0.7284 & 0.8843 & 1.1285 & 1.2438 & 1.4527 & 1.6862 \\              System4.3 & 1.1461 & 1.5331 & 1.7484 & 2.0783 & 2.4182 & 2.6522 \\   
\hline
\multicolumn{6}{c}{$c_2$} \\ 
\hhline{=======}
Version & 5~ns & 6~ns & 7~ns & 8~ns & 9~ns & 10~ns \\
\hline                                                                          System1.13 & 1.125 & 1.150 & 1.144 & 1.130 & 1.143 & 1.112 \\   
System2.1 & 1.158 & 1.164 & 1.068 & 1.153 & 1.146 & 1.098 \\                    System4.3 & 1.157 & 1.079 & 1.153 & 1.141 & 1.117 & 1.139 \\    
\hline

\end{tabular}
\caption{Effective inverse temperature $\beta$ and the rescaling factor $c_2$ achieved by various annealing schedules of different versions of the D-Wave Advantage2 systems according to Eqs.~\eqref{eq:c_2_theory} and \eqref{eq:theoretical_beta}. The estimations are based on the specifications of the time schedule provided by the D-Wave platform~\cite{dwave-schedule}.} 
\label{table:D-Wave}
\end{table}

Through the rest of the study, for the sake of concreteness, we focus on the linear schedule described by $A(t)=\alpha(\tau-t)$ and $B(t)=1$ (note that this is different from the actual schedules used by the D-Wave platform). Provided that $\tau \gg 1/\sqrt{\alpha}$, the above two equations can be approximated as
\begin{align} \label{eq:linear_c2beta}
c_2 &\simeq \frac{\int_0^\infty dt \sin (\alpha n t^2)}{\int_0^\infty dt \sin (\alpha t^2)}=\frac{1}{\sqrt{2}}\;, \nonumber\\
\beta &\simeq 2\int_0^\infty dt \sin(\alpha t^2)=\sqrt{\frac{\pi}{2\alpha}}\;.
\end{align}
We note that Eq.~\eqref{eq:PB} is a good approximation of the Boltzmann distribution for the energy levels in the interval $[-E_*,\,E_*]$ when $\beta \ll 1/E_*$. According to the above formula, this is equivalent to the condition $\alpha \gg E_*^2$. This verifies that QA must be fast enough to ensure the Boltzmann sampling---hence the name ``diabatic quantum annealing''.

\section{Boltzmann sampling\\of the Ising spin systems} \label{sec:sampling_ferro}

Now that we have a concrete recipe for sampling the high-temperature Boltzmann statistics via QA, we examine the performance of the method for some basic spin systems.

\subsection{Infinite-range Ising model}

\begin{figure*}
\includegraphics[width=\textwidth]{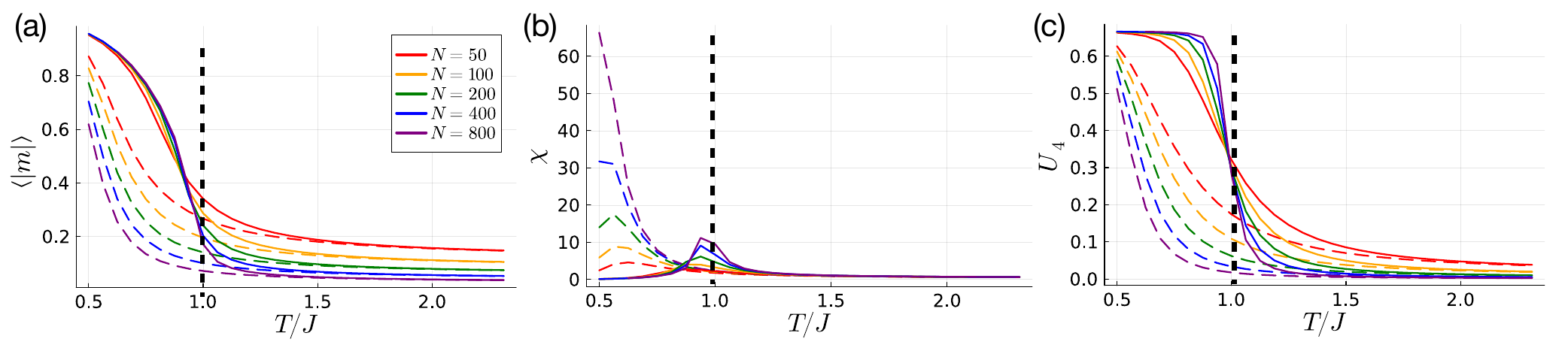}
\caption{\label{fig:observables} Key observables of the ferromagnetic infinite-range Ising model, including (a) magnetization, (b) susceptibility, and (c) Binder cumulant. Solid (dashed) lines represent the theoretical (DQA-estimated) values of the observables. Vertical dashed lines represent the exact boundary between the ferromagnetic and the paramagnetic phases.}
\end{figure*}

We first discuss the case of ferromagnetic Ising models. The ferromagnetic infinite-range Ising model is described by the energy function
\begin{align}
E_\bs = -h \sum_{1 \le i \le N}s_i -\frac{J}{N} \sum_{1\le i<j \le N} s_is_j\;,
\end{align}
where each spin is given by $s_i = \pm 1$, and the factor $1/N$ is needed to ensure that the energy is extensive. The equilibrium free energy of this model can be calculated analytically, with the critical temperature given by $T_\mathrm{c} = 1/\beta_\mathrm{c} = J$ (we use the unit system in which the Boltzmann constant becomes unity). As discussed in Sec.~\ref{sec:DQA}, the Boltzmann distribution of this model can be obtained via DQA that uses the time-dependent Hamiltonian shown in Eq.~\eqref{eq:diabatic}, with
\begin{align}
\h_z = -h \sum_{1 \le i \le N}\hat{\sigma}_i^z -\frac{J}{\sqrt{2}N} \sum_{1\le i<j \le N} \hat{\sigma}_i^z \hat{\sigma}_j^z\;.
\end{align}
While this Hamiltonian has $2^N$ energy eigenstates, all eigenstates associated with the same magnetization have the same energy eigenvalue, so they evolve exactly in the same way during the annealing process. For this reason, the system has only $N$ groups of eigenstates that evolve differently, allowing us to simulate the annealing process for a broad range of $N$.

In Fig.~\ref{fig:kld}, we show how closely the projected distribution $P_\mathrm{Q}$ achieved by DQA with $\alpha$ determined by Eq.~\eqref{eq:linear_c2beta} reproduces the Boltzmann distribution $P_\mathrm{B}$ by plotting the Kullback--Leibler (KL) divergence (or relative entropy)
\begin{align} \label{eq:KLD}
D_\mathrm{KL}(P_\mathrm{B}\|P_\mathrm{Q})=\sum_{\bs} P_\mathrm{B}(\bs)\ln \frac{P_\mathrm{B}(\bs)}{P_\mathrm{Q}(\bs)}
\end{align}
against temperature $T$. In practice, $P_\mathrm{Q}$ is obtained by numerically integrating the Schr\"{o}dinger equation starting from $A(0) = 20$, $B(t) = 20$, with the discretized time step given by $dt = \sqrt{0.1/[A(0)\tau]}$. The same setting is also used for the other representative systems.

The case without the magnetic field ($h = 0$) is examined in Fig.~\ref{fig:kld}(a). As can be expected from the high temperature assumption underlying our method, in the paramagnetic phase ($T > T_\mathrm{c}$), $D_\mathrm{KL}(P_\mathrm{B}\|P_\mathrm{Q})$ increases monotonically as we decrease $T$ and $\alpha$. As $T$ and $\alpha$ decrease further, $D_\mathrm{KL}(P_\mathrm{B}\|P_\mathrm{Q})$ reaches a maximum slightly below $T_c$, and then decreases again as $T$ and $\alpha$ approach $0$ despite the breakdown of the underlying assumptions. The latter phenomenon is due to the concentration of both $P_\mathrm{Q}$ and $P_\mathrm{B}$ around the lowest energy states, which is guaranteed by the adiabatic theorem.

Notably, in the paramagnetic phase, $D_\mathrm{KL}(P_\mathrm{B}\|P_\mathrm{Q})$ seems to converge to a finite value as we increase the system size $N$. The convergence is corroborated by the star symbols ($\bigstar$)---each of these marks the temperature below which $D_\mathrm{KL}(P_\mathrm{B}\|P_\mathrm{Q})$ achieved at a given value of $N$ becomes $30\%$ higher than that achieved when the system size is $N/2$. As $N$ goes to infinity, we expect the $T$ values corresponding to these symbols to converge to $T_\mathrm{c}$, demonstrating that $D_\mathrm{KL}(P_\mathrm{B}\|P_\mathrm{Q})$ saturates to a finite value whenever $T > T_\mathrm{c}$. This is due to the typical energy of the system scaling as $E_* \sim N^0$ in the paramagnetic phase, which means that the assumption $\alpha \gg E_*^2$ required for the validity of our method does not get worse as we increase $N$. In contrast, $D_\mathrm{KL}(P_\mathrm{B}\|P_\mathrm{Q})$ tends to increase with $N$ in the ferromagnetic phase ($T < T_\mathrm{c}$), reflecting that the typical energy of the system scales as $E_* \sim N$ there. In this regime, the assumption $\alpha \gg E_*^2$ continues to worsen as we increase $N$, so the deviations of $P_\mathrm{Q}$ from $P_\mathrm{B}$ can grow uncontrollably in the thermodynamic limit.

The case with the magnetic field ($h \neq 0$), shown in Fig.~\ref{fig:kld}(b), gives us a similar picture. As shown in the main figure, when the system is in the paramagnetic phase ($T > T_\mathrm{c}$) and the magnetic field scales as $h \sim N^{-1/2}$, the typical energy of the system scales as $E_* \sim N^0$. In this case, even as we increase $N$, the assumption $\alpha \gg E_*^2$ does not become worse, resulting in the convergence of $D_\mathrm{KL}(P_\mathrm{B}\|P_\mathrm{Q})$ in the limit $N\to\infty$. On the contrary, if we fix the value of $h$ regardless of $N$, the typical energy of the system satisfies $E_* \sim N$. Then the assumption $\alpha \gg E_*^2$ worsens as we increase $N$, making $D_\mathrm{KL}(P_\mathrm{B}\|P_\mathrm{Q})$ increase with $N$ as shown in the inset. We note that, in both cases, $D_\mathrm{KL}(P_\mathrm{B}\|P_\mathrm{Q})$ is much smaller if the two-body interaction terms in the Hamiltonian are rescaled by a factor of $1/\sqrt{2}$ as specified in the recipe of Sec.~\ref{sec:DQA}.

To demonstrate how the deviations of $P_\mathrm{Q}$ from $P_\mathrm{B}$ affect the observables commonly used in the analyses of phase transitions, in Fig.~\ref{fig:observables} we compare various moments of the magnetization $m \equiv \frac{1}{N} \sum_{i=1}^N s_i$ obtained by $P_\mathrm{B}$ (solid curves) with those obtained by $P_\mathrm{Q}$ (dashed curves) for $h = 0$. Namely, we examine the absolute magnetization $\langle |m| \rangle$, the susceptibility $\chi = \frac{N}{T}(\langle m^2\rangle - \langle m\rangle^2)$, and the Binder cumulant $U_4 = 1 - \frac{\langle m^4 \rangle}{3\langle m^2 \rangle^2}$. All results show that $P_\mathrm{Q}$ yields reliable estimates of observables only in the high-temperature limit, where the discussions of Sec.~\ref{sec:DQA} are valid, and in the low-temperature limit, which is governed by the adiabatic theorem. We also observe that $P_\mathrm{Q}$ effectively underestimates $T_\mathrm{c}$, with the peaks of $\chi$ and the point where $U_4$ obtained at different values of $N$ cross each other shifted from $J$ to much smaller values. This indicates that increasing the amplitudes of the eigenstates associated with typical spin configurations in the ferromagnetic phase takes much longer than expected by the perturbation theory discussed in Sec.~\ref{sec:DQA}.

\begin{figure*}[htb!]
\includegraphics[width=\textwidth]{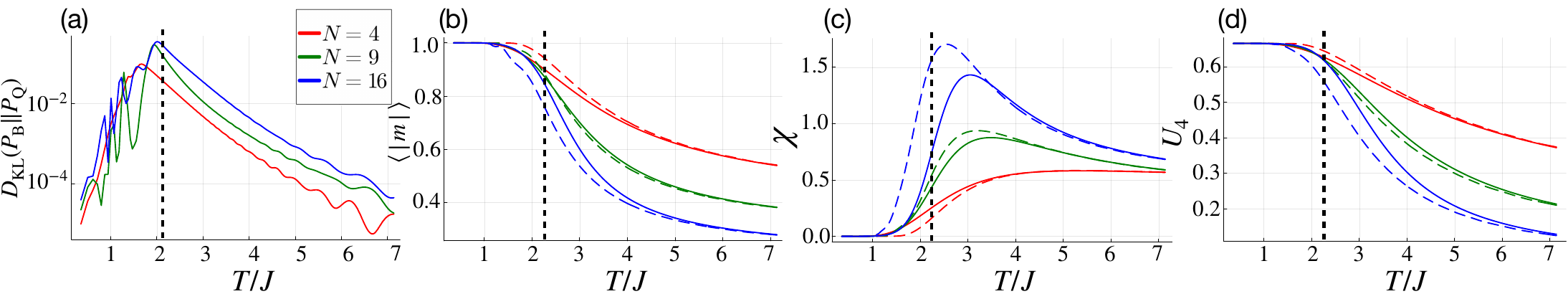}
\caption{\label{fig:2d} Boltzmann sampling of the ferromagnetic Ising model on the square lattice. (a) KL divergence between the theoretical and the DQA-estimated distributions. Comparison between the key observables estimated by the exact Boltzmann distribution (solid lines) and DQA (dashed lines), including (b) magnetization, (c) susceptibility, and (d) Binder cumulant.}
\end{figure*}

\subsection{Ising model on the square lattice}

We also test our method for the ferromagnetic Ising model (with $h = 0$) on the square lattice, whose critical temperature is known to be exactly $T_\mathrm{c} = 1/\beta_\mathrm{c} = \ln(1+\sqrt{2})/2$. In this case, unlike the infinite-range model, the energy of the system is not determined solely by the global magnetization. Instead, the number of eigenstates that evolve differently during the annealing process grows exponentially with $N$, making it practically impossible to check the performance of DQA for very large systems. For this reason, we only used the $N = 2 \times 2$, $3 \times 3$, and $4 \times 4$ square lattices with periodic boundaries.

As shown in Fig.~\ref{fig:2d}(a), the KL divergence of the exact Boltzmann statistics $P_\mathrm{B}$ from $P_\mathrm{Q}$ generated by DQA according to the recipe of Sec.~\ref{sec:DQA} reaches a maximum near the critical temperature $T_c$ (marked by the vertical dashed line), tending to decrease as $T \to \infty$ and as $T \to 0$. The former corresponds to the regime where the perturbation theory of Sec.~\ref{sec:DQA} is valid, and the latter corresponds to the regime governed by the adiabatic theorem. In contrast to the case of the infinite-range Ising model, in this case the KL divergence in the paramagnetic phase does not exhibit any clear sign of convergence to a finite value. Since the Ising model on the square lattice exhibits spatial domains even in the paramagnetic phase, the typical energy of the system always assumes a negative $\mathcal{O}(N)$ value at any finite temperature. This means that Eq.~\eqref{eq:PB} always becomes a poor approximation as $N$ grows to infinity. Thus, for the ferromagnetic two-dimensional Ising model, there is no guarantee that the KL divergence stays $\mathcal{O}(N^0)$ in the paramagnetic phase. 

In Fig.~\ref{fig:2d}(b--d), we also compare $P_\mathrm{B}$ (solid curves) with $P_\mathrm{Q}$ (dashed curves) using the observables $\langle |m| \rangle$, $\chi$, and $U_4$. As observed in the infinite-range model, both distributions yield similar statistics in the low- and the high-temperature limits, and significant deviations are observed in the regime around the critical temperature. Notably, in contrast to the general trend, $P_\mathrm{Q}$ overestimates the transition point for $N = 4$. But as we increase $N$, $P_\mathrm{Q}$ increasingly underestimates $T_\mathrm{c}$, recovering the general trend. While the differences between the two distributions appear to be smaller for this case compared to the case of the infinite-range model, we expect the deviations to become much greater for the annealing processes with larger $N$. 

\subsection{Sherrington–Kirkpatrick model}
Lastly, we apply our method to the Sherrington--Kirkpatrick (SK) model~\cite{Sherrington1975}, whose phase boundaries are exactly known~\cite{Parisi1979,*Parisi1980}. We consider the case where the energy function of the SK model is given by
\begin{align}
E_\bs = \sum_{i<j}J_{ij}s_i s_j,
\end{align}
with $J_{ij}$'s identically and independently distributed according to the normal distribution $\mathcal{N}(0,1/N)$. The system exhibits a critical point $T_c = 1$ separating the spin glass phase ($T < T_c$) from the paramagnetic phase ($T > T_c$). While both phases lack any global order indicated by magnetization, the transition can be detected by examining the overlap parameter $q(\mathbf{s}^1,\mathbf{s}^2)=\frac{1}{N}\sum_{i=1}^N s_i^1s_i^2$, whose expected absolute value $\mean{|q|}$ is nonzero in the thermodynamic limit only in the spin glass phase.

To test the performance of DQA for the SK model, we measured the KL divergence $D_\mathrm{KL}(P_\mathrm{B}\|P_\mathrm{Q})$ between the exact Boltzmann statistics $P_\mathrm{B}$ and the DQA-generated distribution $P_\mathrm{Q}$ for the SK model with $N = 4$, $9$, and $16$, generating $15$ independent realizations of $J_{ij}$ for each $N$. We also estimated $\mean{|q|}$ using DQA and compared the results to the values obtained from the exact Boltzmann statistics.

\begin{figure}[t]
\begin{center}
\includegraphics[width=1\hsize]{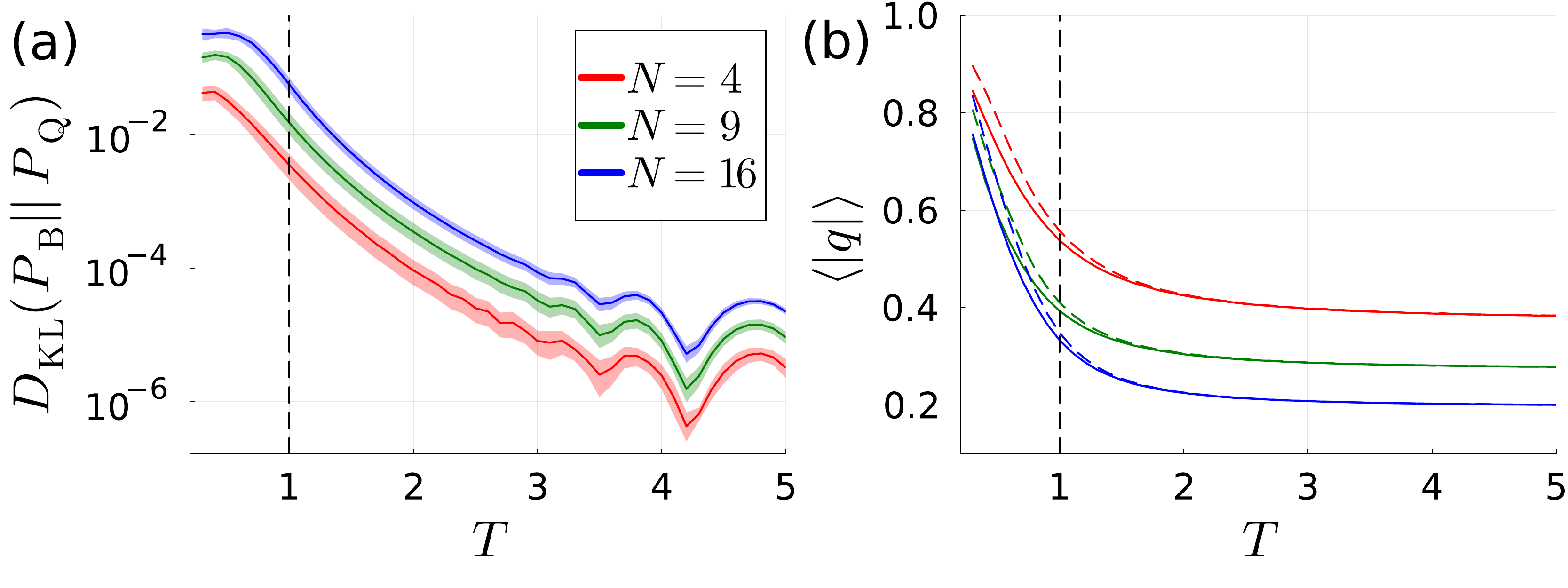}
\caption
{\label{fig:SK-model} Boltzmann sampling of the Sherrington--Kirkpatrick model. (a) KL divergence between the theoretical and the DQA-estimated distributions. (b) Comparison between the overlap parameter estimated by the Boltzmann distribution (solid lines) and DQA (dashed lines).
}
\end{center}
\end{figure}

In Fig.~\ref{fig:SK-model}(a), we plot the mean (solid lines) of $D_\mathrm{KL}(P_\mathrm{B}\|P_\mathrm{Q})$ together with the 90\% confidence intervals (ribbons) for different values of $N$. As can be expected from our previous discussions, we observe that the quantity is smaller in the paramagnetic phase, tending to decrease monotonically with $T$ until it becomes small enough to be dominated by sampling error. We also observe that the KL divergence tends to increase with $N$. Due to the effects of frustration, it is known that the typical energy of the SK model is $\mathcal{O}(N)$ even in the paramagnetic phase despite lacking spatial domains. For this reason, in contrast to the case of the infinite-range Ising model, we expect that the KL divergence of the SK model would not saturate to a finite value as $N \to \infty$. We note that the comparison between $\mean{|q|}$ estimated by DQA (dashed lines) and the exact Boltzmann statistics (solid lines) shown in Fig.~\ref{fig:SK-model}(b) also provides consistent results.

Meanwhile, in the spin glass phase ($T<T_c$), the KL divergence forms a plateau as $T$ decreases to zero. This is in sharp contrast to the ferromagnetic Ising model (see Figs.~\ref{fig:kld}(a) and \ref{fig:2d}(a)), for which the divergence exhibits a peak near $T_c$ and falls to zero as $T$ decreases. The difference can be attributed to the free energy landscapes of the ferromagnetic and the spin glass phases. In the former regime, there are only two dominant free energy minima (\textit{i.e.}, fully ordered states) as $T$ decreases to zero, which greatly reduces the difficulty of Boltzmann sampling in the zero-temperature limit. On the other hand, in the latter regime, there are numerous free energy minima separated by large barriers that similarly contribute to the system's statistics even in the zero-temperature limit. Our results show that DQA struggles to keep track of all those minima and assign proper measures to them, which is also indicated by the growing deviations between $\mean{|q|}$ estimated by DQA (dashed lines) and the exact Boltzmann statistics (sold lines) for $T < T_c$, as shown in Fig.~\ref{fig:SK-model}(b).

To sum up, the simulation results confirm that DQA is indeed a reliable Boltzmann sampler in the high-temperature limit where the typical energy of the system is of order $N^0$. In this regime, we can expect that DQA would offer a significant advantage over the conventional sampling methods via fast, parallel dynamics. However, beyond the regime, there is no guarantee that DQA can reliably sample the Boltzmann distribution in the thermodynamic limit. Just like the conventional MCMC methods, performance of the algorithm deteriorates in the presence of critical fluctuations or rugged free energy landscape.

\section{Derivation details} \label{sec:theory}

Here we present a detailed proof of the theorem stated in Sec.~\ref{sec:DQA}. We start with writing $\h_z=\sum_{n} \h_{z,n}$, where
\begin{align}
\h_{z,n}=-\sum_{i_1<\cdots< i_n} J_{i_1 \cdots i_n} \sigma^z_{i_1}\cdots \sigma^z_{i_n}
\end{align}
represents the contributions from $n$-spin interactions. Then, denoting by $\bs$ the $z$-directional spin configuration and by $E_{\bs,n} = \bra{\bs}\h_{z,n}\!\ket{\bs}$ the corresponding energy component associated with $n$-spin interactions, we can also write
\begin{align}
\h_z = \sum_{\bs} E_{\bs}  \pro{\bs}{\bs}\;,
\end{align}
where $E_{\bs} = \sum_n E_{\bs,n}$. As previously noted below Eq.~\eqref{eq:PB}, the form of the energy function implies that
\begin{align} \label{eq:zero_sum}
\sum_\bs E_{\bs,n} = 0
\end{align}
for any choice of $J_{i_1 \cdots i_n}$'s, which is convenient for normalization.



Let us assume the weak coupling limit and treat $\h_z$ as a perturbation. From now on, we use its interaction-picture representation
\begin{align}
    \h_z(t)&=e^{-i\int_t^\tau ds A(s) \h_x}\,\h_z\,e^{i\int_t^\tau ds A(s) \h_x} \nonumber\\
    &=\sum_n e^{-i\int_t^\tau ds A(s) \h_x}\,\h_{z,n}\,e^{i\int_t^\tau ds A(s) \h_x} \nonumber\\
    &\equiv\sum_n \h_{z,n}(t)\;.
\end{align}
Then, using the Dyson series expansion of Eq.~\eqref{eq:unitary_evol}, the final state $\ket{\psi_\mathrm{f}}$ reached after the annealing process can be expressed as
\begin{align}
    \ket{\psi_\mathrm{f}}
    &=\bigg[1-i\int_0^\tau dt\, B(t) \h_z(t)\nonumber\\
    &\!\!\!\!\!+(-i)^2\!\!\int_0^\tau \!\! dt_1\, \int_0^{t_1} \!\! dt_2\, B(t_1) \h_z(t_1)B(t_2)\h_z(t_2)+\cdots \bigg]\nonumber\\
    &\!\!\!\!\!\times e^{iN\int_0^\tau ds  A(s)}\bigotimes_{i=1}^N\ket{X-}_i \;.\label{eq:Dyson_series}
\end{align}
For convenience, let us define $\phi_t \equiv \int_t^\tau ds \, A(s)$. Through the annealing process, $\h_x$ rotates each Pauli operator about the $x$-axis by the same phase $2\phi_t$. Thus, we can rewrite $\h_{z,n}(t)$ as
\begin{align}\label{eq:H_Pn(t)}
    \h_{z,n}(t)&=-\sum_{i_1<\cdots< i_n} J_{i_1 \cdots i_n}\sigma^z_{i_1}(t)\cdots\sigma^z_{i_n}(t)\;,
\end{align}
where $\sigma_{i}^z(t) \equiv \cos(2 \phi_t)\sigma_{i}^z -\sin(2\phi_t)\sigma_{i}^y$.

Plugging Eq.~\eqref{eq:H_Pn(t)} into Eq.~\eqref{eq:Dyson_series} and using the property
\begin{align}
    \sigma^z_i(t)\ket{X-}_i&=\left[\cos(2 \phi_t) \sigma_{i}^z -\sin(2\phi_t)\sigma_{i}^y\right]\ket{X-}_i\nonumber\\
    &=\left[\cos(2 \phi_t) \sigma_{i}^z -i\sin(2\phi_t)\sigma_{i}^z\right]\ket{X-}_i\nonumber\\
    &=e^{-2i\phi_t}\sigma^z_i \ket{X-}_i\;,
\end{align}
the Dyson series expansion can be truncated at the first order as
\begin{align}
    \ket{\psi_\mathrm{f}}= e^{iN\phi_0}&\bigg[1-i\sum_n\,\!\!\int_0^\tau \!\! dt\,\mathrm{e}^{-2ni \phi_t}B(t)\h_{z,n}\nonumber\\
    &\quad+\mathcal{O}(\h_{z,n}^2)\bigg]\bigotimes_{i=1}^N\ket{X-}_i \label{eq:ket_psi_f=}\;.
\end{align}
After projecting this state on each $\mathbf{s}$, we obtain the probability distribution
\begin{align}
    &P_\mathrm{Q}(\bs)=|\mean{\bs|\psi_\mathrm{f}}|^2 \nonumber\\
    &=\frac{1}{2^N}\left[1-2\sum_n E_{\bs, n} \int_0^\tau \!\!dt\,B(t) \sin (2n\phi_t)+\mathcal{O}(E_{\bs,n}^2)\right]\;,
\end{align}
whose normalization is guaranteed by Eq.~\eqref{eq:zero_sum}.

Now, we compare this distribution with the Boltzmann statistics associated with another Hamiltonian $\h'$, whose high-temperature expansion can be written as
\begin{align}
P_\mathrm{B}(\bs)=\frac{1}{Z'}e^{-\beta E_{\bs}'}=\frac{1}{2^N}
\,\left[1-\beta E_{\bs}'+\mathcal{O}(\beta^2 {E'_{\bs}}^{2})\right]
\end{align}
for the energy eigenvalues $E_\bs'$ of $\h'$, provided that $\sum_\bs E_\bs' = 0$. To ensure that the above two distributions are equal to the first order in $E_{\bs,n}$, we require that the two Hamiltonians are related as
\begin{align}
    \h' = \sum_n c_n \h_{z,n}\;,
\end{align}
where each coefficient $c_n$ is chosen to be
\begin{align} \label{eq:cn}
c_n = \frac{\int_0^\tau dt B(t) \sin (2n\int_t^\tau ds A(s))}{\int_0^\tau dt B(t) \sin (2\int_t^\tau ds A(s))}.
\end{align}
Then, by identifying the inverse temperature
\begin{align} \label{eq:theoretical_beta_2}
\beta=2\int_0^\tau dt\, B(t) \sin (2\int_t^\tau ds A(s))\;,
\end{align}
we can guarantee $P_\mathrm{Q} \simeq P_\mathrm{B}$ up to order $E_{\bs,n}$. From the above two relations, it is straightforward to reproduce Eqs.~\eqref{eq:c_2_theory} and \eqref{eq:theoretical_beta}.

It is natural to ask whether the approximation scheme described above can be systematically improved by including higher-order terms in the Dyson series expansion. However, as shown in Appendix~\ref{app:2nd-dyson}, the second-order contribution diverges, precluding any controlled refinement based on perturbation theory.

To gain further insight, we instead analyze the simplest setting of a single-qubit system, for which the final-state distribution can be computed exactly (see Appendix~\ref{app:single-qubit}). This exact solution shows that, under a linear annealing schedule, DQA deviates from the Boltzmann distribution once the system moves beyond the weak-coupling regime. This behavior points to a fundamental limitation of the linear protocol, suggesting that improved performance may require more sophisticated annealing schedules.

\section{Conclusions} \label{sec:conclusion}

In this study, we proposed diabatic quantum annealing (DQA) as an alternative approach to Boltzmann sampling driven by quantum fluctuations. In contrast to the widely studied freeze-out-based method, which relies on thermalization induced by environmental noise, DQA operates through unitary dynamics, allowing precise control of the effective temperature via the annealing rate. Using the infinite-range Ising model, the two-dimensional Ising model, and the Sherrington--Kirkpatrick model, we demonstrated that the sampling error of DQA in the disordered phase remains bounded even in the thermodynamic limit if the typical energy of the system does not differ macroscopically from that of the infinite-temperature ensemble. Consequently, DQA provides reliable estimates of high-temperature statistics, while its accuracy can deteriorate in the regime where the typical energy grows with the system size. This makes DQA complementary to the freeze-out-based approaches, which are effective only at low, uncontrolled temperatures.

Since classical Markov chain Monte Carlo (MCMC) methods do not suffer from critical slowing down or glassy behaviors in the high-temperature regime, one may ask whether DQA offers any advantage there. To examine the issue in more precise terms, let us consider a system with $M$ links. In an MCMC approach, the computational cost required to generate statistically independent samples scales as $M \times \xi^{z}$, where $\xi$ is the correlation length and $z$ is the dynamical critical exponent. In the paramagnetic phase, where DQA is expected to approximate Boltzmann sampling accurately, the correlation length $\xi$ remains finite. Consequently, in this regime the cost of producing independent samples via MCMC scales as $\mathcal{O}(M)$. In contrast, within the DQA framework, each execution of the annealing schedule yields an independent sample. The dominant computational overhead then arises from the programming stage, whose cost depends on how long-range links are realized in the hardware. If long-range couplings must be implemented logically through minor embedding, then the task of finding a suitable embedding is itself NP-complete, and no polynomial-time solution is guaranteed in general~\cite{Lobe2024}. However, if we focus on systems whose long-range interactions coincide with native physical couplings (for example, those available in the Zephyr connectivity of D-Wave hardware), the programming overhead scales as $\mathcal{O}(M)$. Under this restriction, the asymptotic scaling in $M$ is comparable between DQA and MCMC, and any potential computational advantage of DQA would arise from a favorable constant prefactor rather than from improved asymptotic complexity. According to a recent implementation of DQA on the D-Wave Advantage2 machine, the restricted Boltzmann machine aided by DQA can achieve superior performance compared to the conventional counterpart that relies on the serial MCMC method, while being up to $64$ times faster~\cite{Gilhan2025}. Although the MCMC method can easily make up for this gap by its ease of parallelization, we believe that the gap (and the improved machine learning performance) is significant enough to motivate further explorations of DQA as an alternative computational framework for Boltzmann sampling.

The method still has much room for further development. In the high-temperature regime---where the error remains finite even in the thermodynamic limit---accuracy may be improved by exploring more sophisticated annealing schedules beyond the linear one used in this study. The truncation of the Dyson series expansion at first order, which strongly limits the applicability of the approximation schemes used in this study, may be avoided by suitable resummation methods. Implementing DQA on physical hardware such as the D-Wave platform also introduces new challenges, including efficient encoding of the Hamiltonian and mitigating errors from imperfect realization of fields and couplings. In particular, at the algorithmic level, the sampling error depends solely on the ratio between temperature and coupling strength, as illustrated in Figs.~\ref{fig:kld} and \ref{fig:2d}. However, physical implementations may introduce error contributions that depend on individual values of $T$ and $J$, potentially affecting the sampling accuracy. Addressing these challenges may be important directions for future research.

\begin{acknowledgements}
J.-Y.G., G.K., and Y.B. acknowledge the support by the Global-LAMP Program of the National Research Foundation of Korea (NRF) grant funded by the Ministry of Education (No. RS-2023-00301976). H.K. is supported by National Research Foundation of Korea (NRF) of Korea grant funded by the Korea government (MSIT) (Nos.~2023M3K5A109480511 \& 2023M3K5A1094813) and the KIAS individual grant No.~CG085302 at the Korea Institute for Advanced Study. G.K. is currently at the Department of Statistics and Data Science, Yonsei University.
\end{acknowledgements}

\appendix
\section{Second-order terms of the Dyson series}
\label{app:2nd-dyson}
We obtained Eq.~\eqref{eq:linear_c2beta} as approximations for the inverse temperature $\beta$ and the rescaling factor $c_n$ from the first-order term of the Dyson series. But there remains the question of whether we can improve the approximations by considering higher-order terms of the Dyson series. In the following, we explicitly demonstrate that the Dyson series exhibits divergence starting from the second-order term.

For $N=1$, the second-order term of the Dyson series shown in Eq.~\eqref{eq:Dyson_series} is given by
\begin{widetext}
\begin{align}\label{eq:second_order}
    f_2(\tau)&=(-i)^2\int_0^\tau dt_1 \int_0^{t_1} dt_2 \,B(t_1)\h_z(t_1)B(t_2)\h_z(t_2) \ket{X-}\notag\\
    &= -\int_0^\tau dt_1 \int_0^{t_1}dt_2\, B(t_1)B(t_2) [\cos(2\phi_{t_1})\sigma^z - \sin(2\phi_{t_1}) \sigma^y)][\cos(2\phi_{t_2})\sigma^z - \sin(2\phi_{t_2}) \sigma^y)]\ket{X-}\notag\\
    &=-\int_0^\tau dt_1 \int_0^{t_1} dt_2 \, B(t_1)B(t_2)\{[\cos(2\phi_{t_1})\cos(2\phi_{t_2})+\sin(2\phi_{t_1})\sin(2\phi_{t_2})]\hat{I} \notag\\
    &\qquad\qquad\qquad\qquad\qquad\qquad\qquad+i[\cos(2\phi_{t_1})\sin(2\phi_{t_2})-\sin(2\phi_{t_1})\cos(2\phi_{t_2})]\sigma^x\}\ket{X-}\notag\\
    &=-\int_0^\tau dt_1 \int_0^{t_1} dt_2 \, B(t_1)B(t_2)[\cos(2\phi_{t_1}-2\phi_{t_2}) + i\sin(2\phi_{t_1}-2\phi_{t_2})]\ket{X-}\notag\\
    &=-\int_0^\tau dt_1 \int_0^{t_1} dt_2 \,B(t_1)B(t_2)\exp\!\left[2i(\phi_{t_1} -\phi_{t_2})\right]\ket{X-}.
\end{align}
\end{widetext}

The convergence of this expression depends on $B(t)$ and $\phi_t = \int_t^\tau ds A(s)$. For DQA with the linear schedule ($A(t)=\alpha(\tau-t)$ and $B(t)=1$), Eq.~\eqref{eq:second_order} can be rewritten as
\begin{align}\label{eq:second_order_with_AB}
    f_2(\tau) &= -\!\!\int_0^\tau \!dt_1 \!\!\int_0^{t_1}\! dt_2 \exp\!\left[i(\tau-t_1)^2 - i(\tau-t_2)^2\right]\ket{X-}\notag\\
    &=-\int_0^\tau dt \int_0^t ds \exp\!\left[i(s^2-t^2)\right]\ket{X-},
\end{align}
which diverges as $\tau \rightarrow \infty$. Indeed, in quantum field theory, one often encounters the divergence of higher-order terms in the Dyson series, which necessitates the use of renormalization group theory. However, it remains to be studied how such remedies can be applied to the case of DQA.

\section{Single-qubit case}
\label{app:single-qubit}

\begin{figure}[t]
\begin{center}
\includegraphics[width=1\hsize]{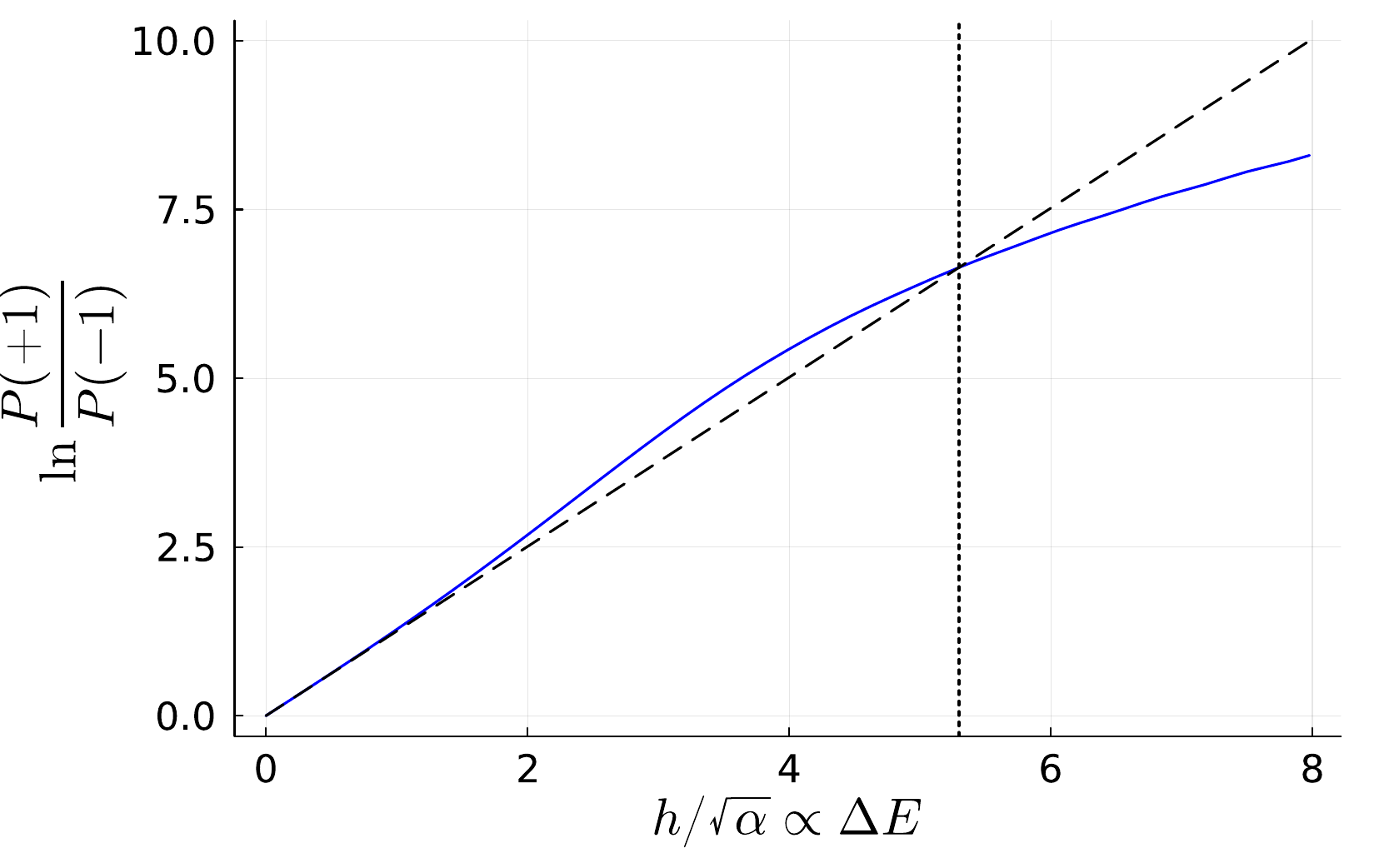}
\caption
{\label{Fig:one_qubit} Performance of DQA as a Boltzmann sampler for the single-qubit case. If DQA samples the Boltzmann distribution exactly, the log ratio between the probabilities of the two energy levels must be a linear function of the energy gap (shown by the dashed line), whose slope is related to the inverse temperature given by Eq.~\eqref{eq:linear_c2beta}. However, an exact solution reveals deviations from the Boltzmann statistics for large $\Delta E$.
}
\end{center}
\end{figure}

Throughout this work, we have relied on the high-temperature expansion (see Eq.~\eqref{eq:PB}) and the Dyson series expansion (see Eq.~\eqref{eq:Dyson_series}) to find an approximate relation (see Eq.~\eqref{eq:linear_c2beta}) between the Boltzmann distribution and the final state of DQA. While the relation is generally applicable to any spin system described by Eq.~\eqref{eq:Es}, its derivation does not provide an explicit demonstration of how the approximation can fail. Fortunately, for a two-level system (which can be constructed from Eq.~\eqref{eq:Es} by setting $N = 1$ and $h \neq 0$) with a linear annealing schedule, the method used to derive the Landau--Zener formula~\cite{Zener1932} allows us to obtain the final state of DQA and assess its proximity to the Boltzmann statistics, as discussed below.

To simplify the notation, we set $A(t)=1$ and $B(t) = -\alpha t$, with the annealing schedule running from $t = -\infty$ to $t = 0$. Note that this schedule is distinct from the annealing schedule $A(t)=\alpha(\tau-t)$ and $B(t)=1$ used in the main text. Since $N=1$ and $h \neq 0$, the Schr\"{o}dinger equation now reads
\begin{align}
    \frac{d}{dt}\ket{\psi} =-i(h\sigma^z -\alpha t \sigma^x)\ket{\psi}.
\end{align}
The amplitudes of $\ket{\psi}$ can be characterized via $\ket{\psi} = \psi_{X+}\ket{X+}+\psi_{X-}\ket{X-}$, where $\ket{X+}$ and $\ket{X-}$ are the eigenstates of $\sigma^x$ with $\sigma^x\ket{X+}=\ket{X+}$ and $\sigma^x\ket{X-}=-\ket{X-}$. Then, the Schr\"{o}dinger equation implies
\begin{align}
    \frac{d}{dt}\psi_{X+} &=  - ih\psi_{X-}+i \alpha t \psi_{X+} \label{eq:time_dif_eq_1},\\
    \frac{d}{dt}\psi_{X-} &= -i\alpha t\psi_{X-} -i h \psi_{X+} \label{eq:time_dif_eq_2},
\end{align}
which can be combined to yield
\[
\frac{d^2}{dt^2}\psi_{X-} +(\alpha^2t^2+i\alpha+h^2)\psi_{X-} =0.\label{eq:second_ordered_ode}
\]
Using the substitutions
\begin{align}
    s = \mathrm{e}^{-i\pi/4}\sqrt{2\alpha} t, \quad
    \nu = \frac{ih^2}{2\alpha}-1,
\end{align}
we can write
\begin{align}
    \frac{d^2}{ds^2}\psi_{X-} +\left(-\frac{s^2}{4}+\frac{1}{2}+\nu \right)\psi_{X-} =0\label{eq:second_ordered_ode_2},
\end{align}
whose general solution is given by the parabolic cylinder function $D_\nu$ as
\[
\psi_{X-}(s) =A_+ D_{\nu}(s) + A_-D_{\nu}(-s).
\]
According to the setting of DQA described in Sec.~\ref{sec:DQA}, the initial condition must satisfy
\begin{align}
    \psi_{X-}(0) &= -\frac{1}{\sqrt{2}},\\
    \dot{\psi}_{X-}(0) &= -ih\psi_{X+}(0) = -\frac{ih}{\sqrt{2}},
\end{align}
which determines the coefficients
\begin{align}
    A_+ &= -\frac{\Gamma(\tfrac{1-\nu}{2})}{2^{\frac{\nu+3}{2}}\sqrt{\pi}}+\frac{ih e^{\frac{i\pi}{4}}}{4\sqrt{\alpha}}\frac{\Gamma(-\tfrac{\nu}{2})}{2^{\frac{\nu+1}{2}}\sqrt{\pi}},\\
    A_- &= -\frac{\Gamma(\tfrac{1-\nu}{2})}{2^{\frac{\nu+3}{2}}\sqrt{\pi}}-\frac{ih e^{\frac{i\pi}{4}}}{4\sqrt{\alpha}}\frac{\Gamma(-\tfrac{\nu}{2})}{2^{\frac{\nu+1}{2}}\sqrt{\pi}}.
\end{align}
Now, we can obtain the probability of $s = -1$ achieved by DQA
\begin{align}
    P_Q(-1)& = \lim_{t\rightarrow -\infty}  \left|\psi_{X-}\left(\mathrm{e}^{-i\pi/4}\sqrt{2\alpha} t\right)\right|^2 \notag\\
    &=\left|\frac{\sqrt{2\pi}}{\Gamma(-\nu)}\mathrm{e}^{-i\pi\nu/4}A_+\right|^2,
\end{align}
from which the probability of the $s = +1$ is trivially obtained as $P_Q(+1) = 1-P_Q(-1)$. Through this derivation, we can now express $P_Q(\pm 1)$ as functions only of $h/\sqrt{\alpha} = \sqrt{-2i(\nu+1)}$.

If the final state of DQA samples the exact Boltzmann distribution, $P_Q(\pm 1)$ obtained above must satisfy $\ln[P_Q(+1)/P_Q(-1)] \propto \Delta E \propto h$, where $\Delta E$ is the energy gap between the two levels in the final state. However, as illustrated by Fig.~\ref{Fig:one_qubit}, the linear relation between $\ln[P_Q(+1)/P_Q(-1)]$ and $\Delta E$ holds only when the energy gap is small. As the energy gap grows, DQA first tends to over-represent the lower energy level; as the energy gap grows even further, DQA eventually under-represents the lower energy level. These results show that, even after the effects of collective spin behaviors are ruled out, there are some fundamental limitations to the capacities of DQA as a Boltzmann sampler.

\bibliography{refs}

\end{document}